\documentclass[10pt,aps,prl,superscriptaddress,twocolumn,floatfix]{revtex4-2}

\usepackage{graphicx}
\usepackage{amsmath}
\usepackage{xcolor}

\newcommand{\coso}{Cu$_2$OSeO$_3$}
\newcommand{\tc}{$T_{\textrm{C}}$}
\newcommand{\trexs}{$t-$REXS}

\begin{document}

\title{Direct observation of the exchange anisotropy in the helimagnetic insulator \coso}

\author{Priya R. Baral}
\email{priya.baral@psi.ch, victor.ukleev@helmholtz-berlin.de}
\affiliation{Crystal Growth Facility, Institute of Physics, \'Ecole Polytechnique F\'ed\'erale de Lausanne (EPFL), CH-1015 Lausanne, Switzerland}
\affiliation{Chair of Computational Condensed Matter Physics, Institute of Physics, \'Ecole Polytechnique F\'ed\'erale de Lausanne (EPFL), CH-1015 Lausanne, Switzerland}
\affiliation{Laboratory for Neutron Scattering and Imaging (LNS), Paul Scherrer Institute (PSI), CH-5232 Villigen, Switzerland}

\author{Oleg I. Utesov}
\affiliation{Center for Theoretical Physics of Complex Systems, Institute for Basic Science, Daejeon 34126, Republic of Korea}

\author{Chen Luo}
\affiliation{Helmholtz-Zentrum Berlin f\"ur Materialien und Energie, D-14109 Berlin, Germany}

\author{Florin Radu}
\affiliation{Helmholtz-Zentrum Berlin f\"ur Materialien und Energie, D-14109 Berlin, Germany}

\author{Arnaud Magrez}
\affiliation{Crystal Growth Facility, Institute of Physics, \'Ecole Polytechnique F\'ed\'erale de Lausanne (EPFL), CH-1015 Lausanne, Switzerland}

\author{Jonathan S. White}
\affiliation{Laboratory for Neutron Scattering and Imaging (LNS), Paul Scherrer Institute (PSI), CH-5232 Villigen, Switzerland}

\author{Victor Ukleev}
\affiliation{Helmholtz-Zentrum Berlin f\"ur Materialien und Energie, D-14109 Berlin, Germany}
\affiliation{Swiss Light Source, Paul Scherrer Institute (PSI), CH-5232 Villigen, Switzerland}


\date{\today}

\begin{abstract}
The helical magnetic structures of cubic chiral systems are well-explained by the competition among Heisenberg exchange, Dzyaloshinskii-Moriya interaction, cubic anisotropy, and anisotropic exchange interaction (AEI). Recently, the role of the latter has been argued theoretically to be crucial for the low-temperature phase diagram of the cubic chiral magnet \coso, which features tilted conical and disordered skyrmion states for a specific orientation of the applied magnetic field ($\mu_0 \vec{\mathrm{H}} \parallel [001]$). In this study, we exploit transmission resonant x-ray scattering (\trexs) in vector magnetic fields to directly quantify the strength of the AEI in \coso, and measure its temperature dependence. We find that the AEI continuously increases below 50\,K, resulting in a conical spiral pitch variation of $10\%$ in the (001) plane. Our results contribute to establishing the interaction space that supports tilted cone and low-temperature  skyrmion state formation, facilitating the goals for both a quantitative description and eventual design of the diverse spiral states existing amongst chiral magnets.
\end{abstract}

\maketitle



In recent years, skyrmions in magnetic materials have attracted significant interest due to their potential spintronic functionalities that promise a paradigm shift in magnetic random access memory, data storage technologies, energy saving, and unconventional computing \cite{fert2017magnetic,everschor2018perspective,song2020skyrmion}. Skyrmions are typically found in thin films with asymmetric interfaces \cite{sampaio2013nucleation} and bulk noncentrosymmetric crystals, such as chiral and polar helimagnets \cite{bogdanov1994thermodynamically,tokura2020magnetic}.

The ground-state helical magnetic structures of cubic chiral systems are well-described by the Bak-Jensen model, which considers the interplay between Heisenberg exchange interaction, Dzyaloshinskii-Moriya interaction (DMI), anisotropic exchange interaction (AEI), and cubic anisotropy (CA) \cite{bak1980theory,nakanishi1980origin,maleyev2006cubic}. The orientation of the helix axis is determined by a subtle interplay among DMI, AEI, and CA. The AEI has been broadly neglected due to its weak impact on experimental observations. However, both cubic and exchange anisotropies play a crucial role in determining the propagation direction of the helix \cite{maleyev2006cubic}, and ultimately, the orientation of any field-induced skyrmion lattice (SkL) in these materials \cite{leonov2020field,adams2018response,kindervater2020evolution}. Moreover, in centrosymmetric materials the competition between AEI and single-ion anisotropy can stabilize SkL even without DMI \cite{hirschberger2021nanometric}.  

\begin{figure*}
\begin{center}
\includegraphics[width=1\linewidth]{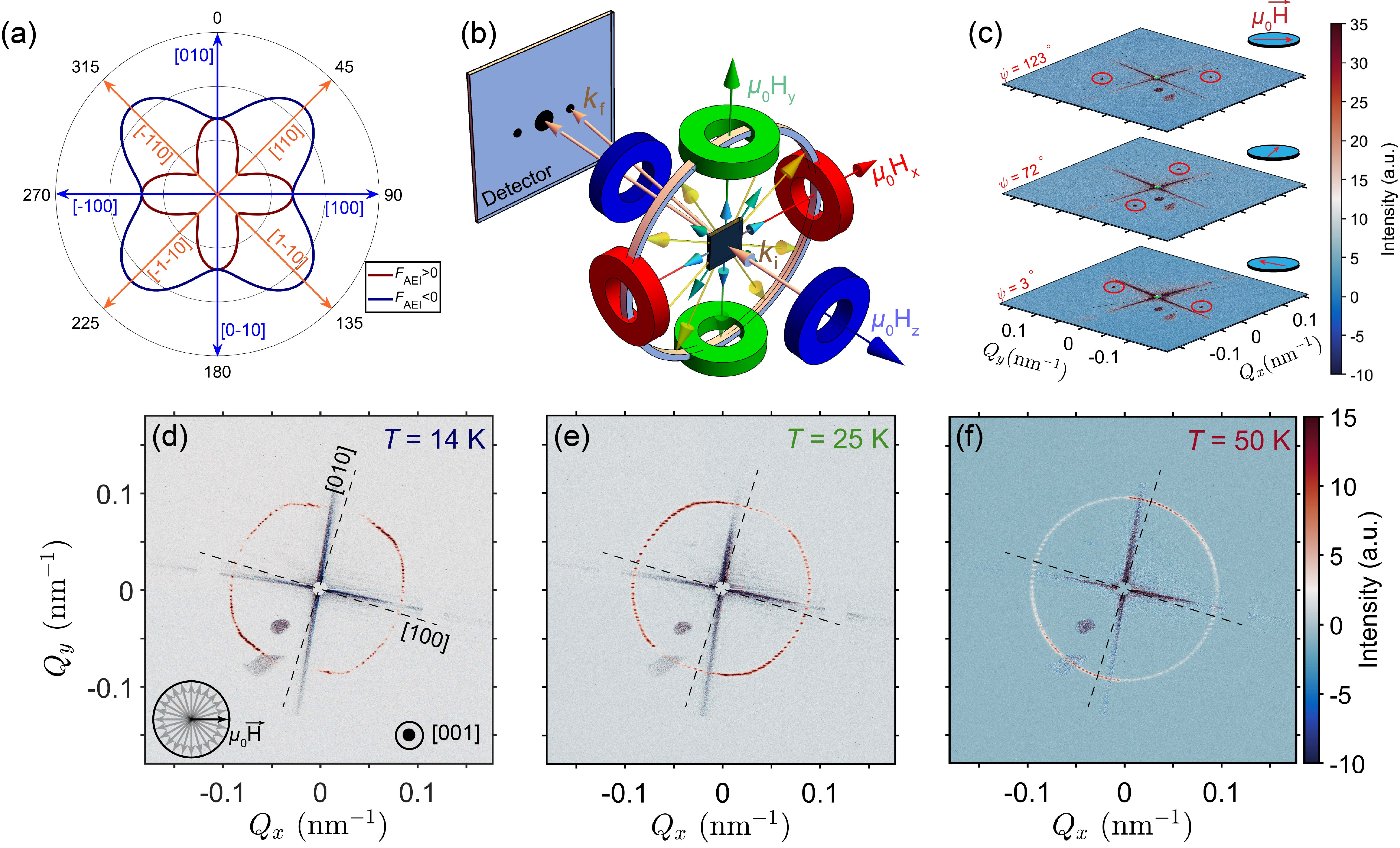}\vspace{3pt}
        \caption{(a) Illustration of the spiral modulation vector $Q$ dependence on azimuthal angle in (110) plane for positive and negative signs of the exchange anisotropy constant, $F_\textrm{AEI}$. (b) Sketch of the geometry of the \trexs~experiment at VEKMAG \cite{noll2016mechanics}. The magnetic field was vectorially varied in $x-y$ plane. (c) \trexs~patterns measured in conical states for different azimuthal angles, $\psi=3^\circ,72^\circ,123^\circ$ at $T=14$\,K. Sum of the \trexs~patterns over all measured $\psi$ angles from 0 to 180$^\circ$ at (d) 14\,K, (e) 25 K and (f) 50\,K.}
        \label{fig1}
\end{center}
\end{figure*}

\begin{figure*}
\begin{center}
\includegraphics[width=1\linewidth]{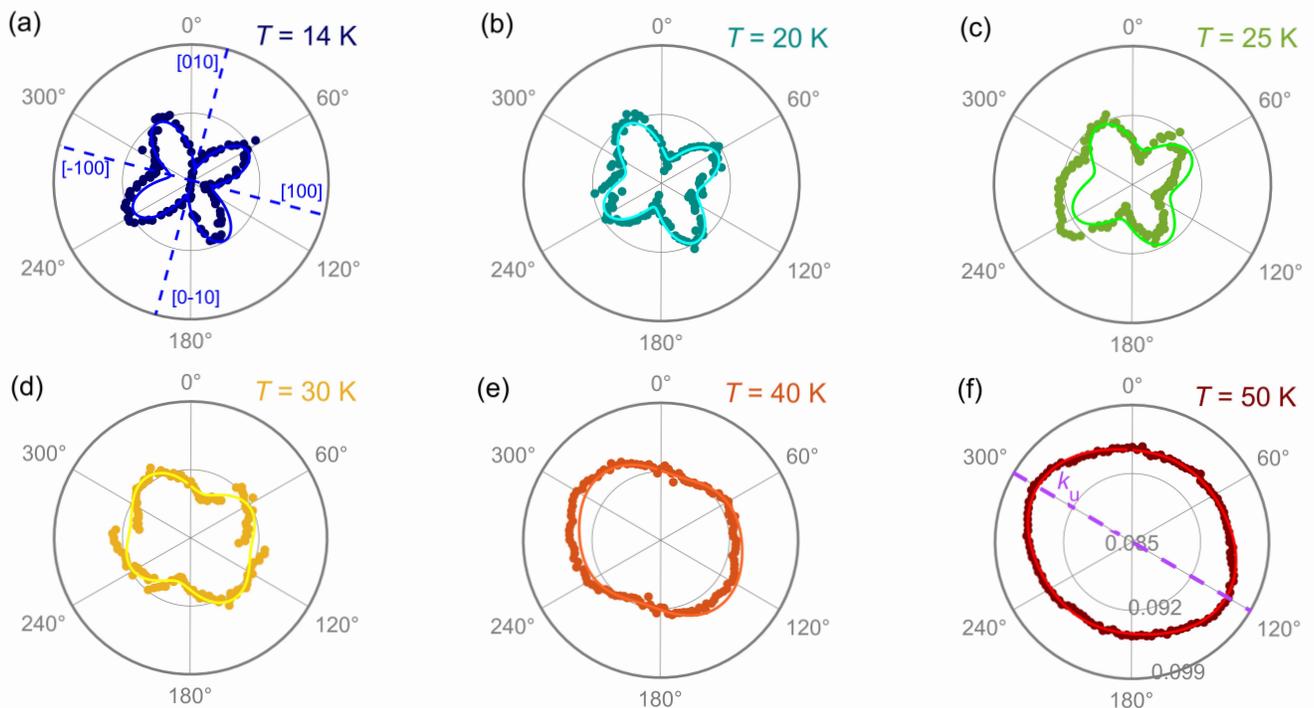}\vspace{3pt}
        \caption{Polar plots of the extracted conical spiral wavevector $Q$ as a function of angle $\psi$ at (a) 14\,K, (b) 20\,K, (c) 25\,K, (d) 30\,K (e) 40\,K and (f) 50\,K. Solid lines correspond to the fit according to the Eq. \ref{eq1} including the offset of $17^\circ$ between $\psi=0^\circ$ and [100] axis due to imperfect sample mounting (see Supplementary information for more details on the sample orientation \cite{supp}). The radial scale for $Q$ is given in the panel (f) and is the same for all panels (a--f).}
        \label{fig2}
\end{center}
\end{figure*}

In cubic chiral magnets, the nontrivial temperature evolution of anisotropic interactions has been demonstrated in $B20$s \cite{lebech1989magnetic,grigoriev2009crystal}, $\beta$-Mn alloys \cite{preissinger2021vital,karube2020metastable} and Zn-doped \coso \cite{moody2021experimental}. Often, the unambiguous experimental distinction between the effects of cubic and exchange anisotropies is challenging since they both affect macroscopic parameters, such as the transition fields between helical and conical states, and conical and field-polarized states \cite{maleyev2006cubic,grigoriev2015spiral}. Even neutron scattering techniques that are sensitive to microscopic material's parameters, are often unable to discriminate these two interactions without an additional theoretical model. According to phenomenological models, a fixed sign of the cubic anisotropy constant $K_c>0$, ground-state helical spirals in cubic chiral magnets propagate along $[ 100 ]$-axes in the case of a positive AEI constant $F_\textrm{AEI}>0$ (e.g. in Fe$_{0.85}$Co$_{0.15}$Si \cite{grigoriev2006magnetic}) and along $[ 111 ]$-axes if $F_\textrm{AEI} < 0$ (e.g. in MnSi \cite{grigoriev2009crystal}).

Notable examples in earlier work where the role of the AEI shows up clearly include in FeGe, where the reorientation of the spiral propagation vector from $[100]$ to $[111]$ due to a sign change of the AEI  \cite{lebech1989magnetic}, and in Zn-doped \coso~where a sign change of the AEI is also argued, albeit with no reorientation of the helix due to the predominance of CA \cite{moody2021experimental}.

Here, we focus on pristine \coso, a magnetoelectric chiral magnet with \tc=58\,K \cite{seki2012observation,white2014electric} which, in addition to conventional helical, conical, and SkL phases, also features several exotic metastable states, such as square and elongated SkL phases \cite{takagi2020particle,aqeel2021microwave}. Recently, the competition between the cubic and exchange anisotropies was argued to be crucial for the manifestation of unusual yet thermodynamically stable magnetic phases in \coso: tilted conical spiral and disordered skyrmions that emerge at low temperatures when a magnetic field is applied along one of the cubic axes \cite{qian2018new,chacon2018observation,bannenberg2019multiple}. Due to the magnetoelectric coupling of \coso~\cite{white2014electric}, its versatile magnetic phase diagram, and the ability to train the low-temperature skyrmion phase \cite{aqeel2021microwave} the material is particularly interesting for exploring chiral magnet based applications paradigms \cite{Lee2022task}. Furthermore, developing the understanding of the fundamental mechanism of skyrmion stabilization through anisotropy engineering paves the way for magnetic phase manipulation amongst the known skyrmion hosts, and which can be particularly relevant for room-temperature topological magnetic textures among noncentrosymmetric materials with high magnetic ordering temperatures such as chiral $\beta$-Mn-type alloys (\tc\, up to 400\,K) \cite{karube2020metastable} and LiFe$_5$O$_8$ (\tc$\sim$900\,K) \cite{iguchi2015nonreciprocal}. Therefore, the unambiguous microscopic, quantitative determination of anisotropic interactions such as the AEI in model chiral magnets such as pristine \coso is highly desirable.

Here we exploit the high momentum-space resolution of transmission resonant elastic x-ray scattering ($t$-REXS) in vector magnetic fields \cite{ukleev2021signature} to quantify directly the AEI in \coso. We obtain the following key results. First, the angular variation of the conical spiral pitch in the (100) plane observed by $t$-REXS agrees with a theory allowing the quantitative extraction of the AEI. Second, in contrast to both FeGe \cite{lebech1989magnetic} and lightly Zn-doped \coso~\cite{moody2021experimental}, the sign of the AEI always remains negative across the entire temperature range below $T_c$. Third, the magnitude of the AEI increases continuously below 50\,K, correlating with the stability window of the tilted cone and disordered skyrmion phases. Taken together, our results implicate the thermal evolution of the AEI and its competition with CA as determining the structure of the phase diagram, and contribute quantitatively towards the foundation of the theoretical modeling and manipulation of spin textures in \coso~and other anisotropic chiral magnets.

In the isotropic case, the spiral propagation vector is proportional to the ratio of the DMI and exchange, $Q_0 \sim D/J$. When the anisotropic interactions come into play, the conical structure becomes distorted and, in general, contains infinite number of harmonics, and the exact solution for the spin structure can hardly be found. However, if the characteristic helical energy is much larger than the anisotropic contributions one can use a perturbative approach. In our case, in order to obtain corrections to the spiral vector, we obtain approximate solution for the sine-Gordon equation describing in-plane magnetization component with the anisotropy-induced terms. The latter are due to AEI, CA and easy-plane anisotropy originating from the tensile strain of the lamella. Importantly, the leading order approximation allows us not to take into account small local variations of the conical angle. Details of the derivation of the following equation are given in the Supplementary Information~\cite{supp}. 

At high temperatures the cubic anisotropy is small~\cite{grigoriev2022transition} (it is of the fourth order in the magnetization modulus) and we consider only the effect of AEI and easy-plane. The result for the spiral vector reads
\begin{eqnarray} \label{eq1}
Q  &=& Q_0 \Biggl\{  1 - \dfrac{F_{\textrm{AEI}}\sin^{2}{2\psi}}{4J}\Biggr\} - \dfrac{JZ^{2}\cot^{2}\alpha}{2D^{3}}\sin^{2}2(\psi-\phi) \nonumber \\ &&
 - \dfrac{JZ^{2}}{8D^{3}}\sin^{4}(\psi-\phi). 
\end{eqnarray}
Here $\psi$ is the azimuth angle of the conical helicoid propagation vector in the $(001)$ plane, $\alpha$ is the conical angle ($\alpha=0$ in the fully polarized phase), $Z$ is the easy-plane anisotropy constant and $\phi$ indicates the corresponding axis direction. At small temperatures the AEI-induced correction in the first term of Eq.~\eqref{eq1} dominates; other terms, including the one stemming from CA (see~\cite{supp}), being less prominent. In addition, we have tried to fit the experimental data considering the higher-order exchange anisotropy term, and found that the result is the same within the error bar. Therefore, it is excluded from the analysis.

A polar plot of the spiral wavevector $Q$ depending on its orientation in the $(001)$ is shown in Fig. \ref{fig1}a for positive and negative $F_\textrm{AEI}$ constants as an example. For $F_\textrm{AEI}>0$ the propagation vector of the spiral in the ground state is favored by AEI along $\langle001\rangle$, while for negative $F_\textrm{AEI}$ spirals propagate along diagonals of the cubic lattice $\langle111\rangle$. Experimentally, the conical propagation vector can be determined at will azimuthally in the (001) crystal plane by a finite vector magnetic field (Fig. \ref{fig1}b). The detailed description of the samples and the measurement setup is given in the Supplementary Information \cite{supp}. Each conical state was prepared using the following procedure. First, the helical state was achieved by cooling the sample down to the target temperature at zero field ($T = 14$\,K), followed by ramping up the magnetic field to $70$\,mT to force the sample into the field-polarized state. The field was then ramped down to $30$\,mT in order to remain inside the conical phase for a particular $\psi$. After each acquisition, the magnetic field was again ramped up to $70$\,mT, followed by changing its in-plane direction. This protocol was repeated for each $\psi$ ranging between 0 and 180$^\circ$ with a 3$^\circ$ step.

At each particular sample temperature, the intensity corresponding to each of the Friedel pair of conical peaks for the measured $\psi$ was extracted and summed up. The resulting patterns measured at the lowest (14\,K), intermediate (25\,K) and highest (50\,K) temperatures are shown in Figs. \ref{fig1}d--f. At $T = 50$\,K, the intensity profile appears almost circular, but with a slight elliptical distortion (\ref{fig1}f). The observed ellipticity in the intensity profile is an indication of the uniaxial anisotropy induced by strain arising from the contacts made on the lamella sample during FIB milling \cite{shibata2015large,okamura2017directional,ukleev2020metastable}. As the temperature decreases, the profile starts to develop subtle features along the marked crystallographic axes. On cooling, the azimuthal scattering intensity distribution profile deviates from the ellipticity seen at 50\,K and develops extra humps along the in-plane $[110]$ directions. This is most strongly pronounced at the base temperature of 14\,K, as shown in (\ref{fig1}d). Concomitantly, $|Q|$ along the $[100]$ directions is found to be the minimum, and $Q || [110]$ the maximum. Interestingly, in contrast to FeGe \cite{ukleev2021signature}, the helical spiral was observed to always revert the orientation of propagation back to the $[100]$ direction upon leaving the in-plane conical phase through reduction of the field. This is another manifestation of the strong CA in \coso.

\begin{figure}
\includegraphics[width=1\linewidth]{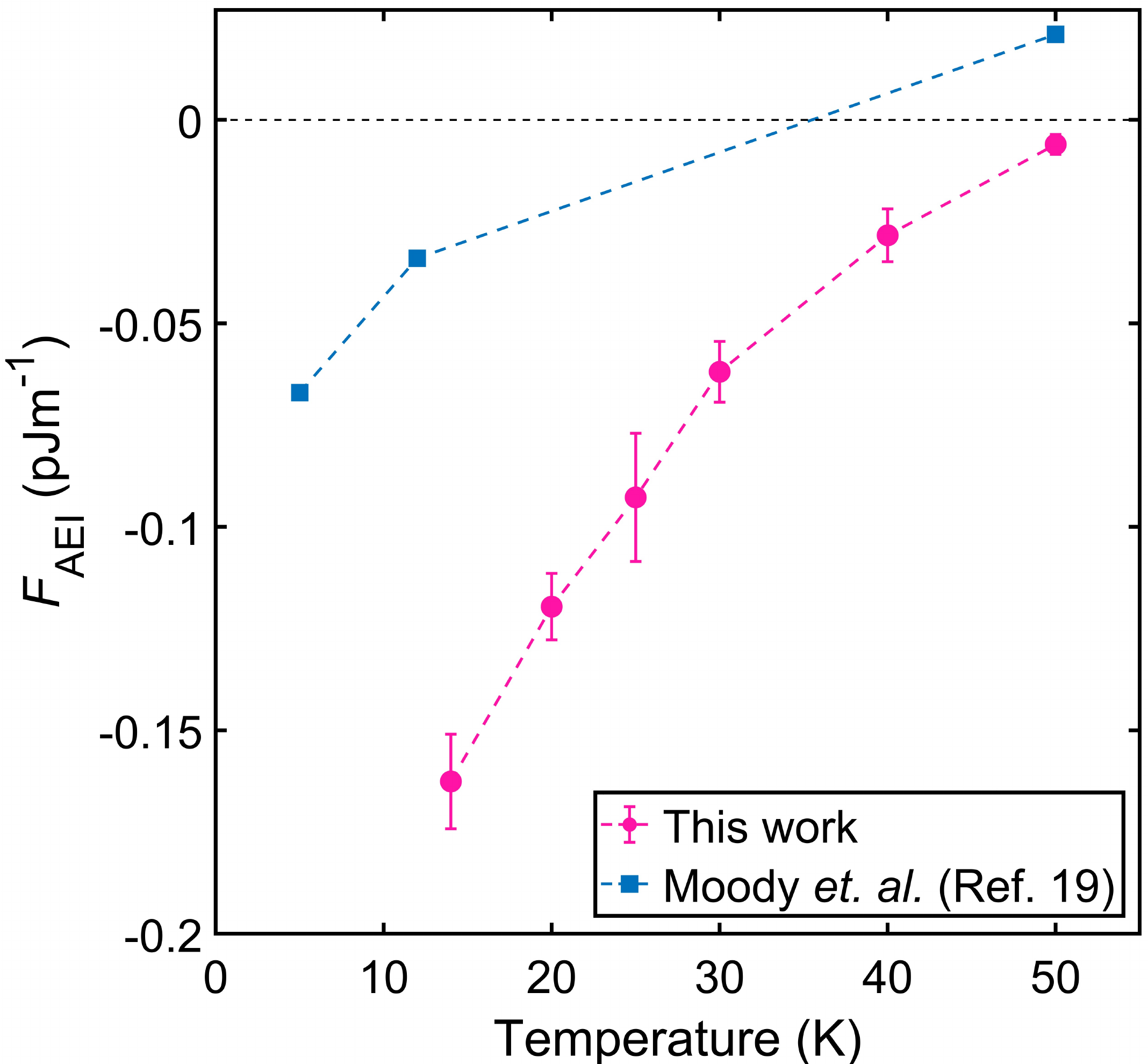}\vspace{3pt}
        \caption{Temperature dependence of the exchange anisotropy constant $F_\textrm{AEI}$ extracted from the fit of $Q(\psi)$ according to Eq. \ref{eq1}. The black dashed line is a guide to the eye.}
        \label{fig3}
\end{figure}

In the next step, both the radial and azimuthal profiles of the diffracted intensity at each $\psi$ were examined. In order to only contain a single Bragg peak, a sector box of 3$^\circ$ angular width was chosen around each. Also, both peaks from the Friedel pair were analyzed separately, using mirror sectors, providing us with information on $|Q|$ in all four quadrants simultaneously. Polar plots of the extracted peak position $Q_\textrm{c}(\psi)$ in Figs. \ref{fig2}a--f directly show the anisotropic nature of the conical spirals in \coso, and how this develops on cooling. The direct influence of temperature dependence of the AEI on $|Q_\textrm{c}|$ can be seen clearly in Fig. \ref{fig2}. At the lowest $T=14$\,K (Fig. \ref{fig2}a), $Q_\textrm{c}$ varies between 0.086~nm$^{-1}$ along [100] to 0.092~nm$^{-1}$ along [110] in a monotonous fashion, as it expected according to Eq. \ref{eq1}. This shows that in \coso~the AEI is most pronounced at low temperatures, resulting in the conical spiral pitch variation up to 10\% between the conical spirals oriented along [100] and [110].

In order to quantify the AEI constant, $Q(\psi)$ dependencies were fitted according to Eq.~\eqref{eq1} (solid lines in Fig. \ref{fig2}). The result is shown in Fig. \ref{fig3}, where $|F_\textrm{AEI}|$ clearly tends to monotonically increase towards low temperatures and reaches of $F_\textrm{AEI}=-0.163\pm0.012$ pJm$^{-1}$ at 14\,K, and practically vanishes at 50\,K. The strain-induced anisotropy terms containing $Z$ in Eq. \ref{eq1} do not show significant variation as a function of temperature (see Supplementary Information \cite{supp}).

The sign of the AEI constant $F_\textrm{AEI}$ in \coso~is negative in the whole temperature range, in contrast to previous results on Zn-doped \coso~(Fig. \ref{fig3})  \cite{moody2021experimental}. As shown before, a few percent Zn doping can modify the microscopic properties of pristine \coso~significantly \cite{vstefanvcivc2018origin}, and our data supports this conclusion. Moreover, this suggests that one can tune the microscopic parameter with a small doping, and hence finely tailor the helical (skyrmion lattice) pitch and stability windows of anisotropy-driven phases. Importantly, at low temperatures the two systems consistently demonstrate the strong contribution of the magnetocrystalline anisotropy that pins the spiral wavevector along $[100]$. Nonetheless, the competition between the AEI and cubic anisotropies favoring different orientation of magnetic spirals is known to stabilise more unusual magnetic spiral superstructures such as tilted conical and disordered skyrmion phases \cite{qian2018new,bannenberg2019multiple,leonov2020field}. A fine balance between  AEI and CA is required to theoretically reproduce low-temperature magnetic phases in \coso~\cite{bannenberg2019multiple}. The strong enhancement of the AEI at low temperatures is evident from our data and provides a much needed quantitative basis for the stability of tilted conical and disordered skyrmion states proposed by the theory. Therefore, a chemical tuning of the AEI would be a promising approach to stabilize new phases far below \tc~in other known cubic chiral magnets.

In summary, the study of the anisotropic exchange interaction (AEI) in the cubic chiral magnet \coso~using transmission resonant x-ray scattering in vector magnetic fields has revealed that the sign of the AEI energy constant is negative in the whole temperature range below $T_\textrm{C}$, and continuously increases below 50\,K, to reach $F_\textrm{AEI}= -0.163$\,pJm$^{-1}$ at our lowest temperature of 14\,K. The sign of the AEI constant is negative in the whole temperature range, pointing to a stronger contribution of cubic anisotropy that pins the spiral propagation vector along $[001]$. The magnitude of $F_\textrm{AEI}$ is of the same order as in FeGe but with an opposite sign. Our measurements of the strong enhancement of the AEI at low temperatures provide a quantitative basis for phenomenological theories that describe how competing anisotropies in chiral magnets can stabilize novel complex spiral magnetic states such as tilted conical and disordered skyrmion phases. Additionally, we have presented a theoretical and experimental framework for quantifying AEI in cubic chiral magnets, and distinct it from CA, which is valuable for comparison with $ab-initio$ theories and for understanding the role of AEI in the emergence of skyrmions and other exotic magnetic state. Similar approach can be further developed for broader class of anisotropic magnets with long-periodic spin modulations stabilized by other mechanisms, such as frustrated interactions \cite{ballou1987helimagnetism,yu2012magnetic,lin2016ginzburg}.  

\section*{Acknowledgements}

Authors thank A. Leonov for fruitful discussions, E. Deckardt, M. Bednarzik and Th. Jung for their help in preparation of the membranes at PSI, B. Bartova for the assistance in FIB nano-fabrication at EPFL CIME, and K. Schwarzburg for the help with the scanning electron microscopy measurement in the Corelab Correlative Microscopy and Spectroscopy at Helmholtz-Zentrum Berlin. The \trexs~ experiment was carried out at the beamline PM-2 VEKMAG at BESSY II synchrotron as a part of the proposal 212-10682 ST. P.R.B., J.S.W., A.M., V.U. acknowledge funding from the SNSF Project Sinergia CRSII5\_171003 NanoSkyrmionics. P.R.B. also acknowledges SNSF grant no. 200020\_182536 (Frustration in structures and dynamics). We acknowledge financial support for the VEKMAG project and for the PM2-VEKMAG beamline by the German Federal Ministry for Education and Research (BMBF 05K2010, 05K2013, 05K2016, 05K2019) and by HZB. F.R. acknowledge funding by the German Research Foundation via Project No. SPP2137/RA 3570. V.U. thanks the SNSF National Centers of Competence in Research in Molecular Ultrafast Science and Technology (NCCR MUST-No. 51NF40-183615) for the financial support. 




\bibliography{biblio}

\end{document}